\documentclass[preprints,article,accept,oneauthor]{Definitions/mdpi}

\firstpage{1}
\pubvolume{1}
\issuenum{1}
\articlenumber{0}
\pubyear{2026}
\copyrightyear{2026}
\datereceived{}
\daterevised{}
\dateaccepted{}
\datepublished{}
\usepackage{graphicx}

\newcommand{\avg}[1]{\left\langle #1 \right\rangle}
\newcommand{\dd}{\mathrm{d}}

\newcommand{\YN}{Y_{\Lambda}}
\newcommand{\rhoB}{\rho_{B}}
\newcommand{\rhoL}{\rho_{\Lambda}}

\newcommand{\VTBF}{V_{NN}^{(3),\mathrm{eff}}}

\Title{Hyperonic Softening versus Nucleonic Three-Body Repulsion in Hypernuclear Matter within a microscopic approach}
\nolinenumbers
\Author{Mahboubeh Shahrbaf $^{1,2,*}$}
\address{%
$^{1}$ \quad Institute of Theoretical Physics, University of Wroclaw,
Plac Maxa Borna 9, 50-204 Wroclaw, Poland; \\
$^{2}$ \quad Nishina Center, RIKEN, Hirosawa 2-1, Wako, 351-0198, Saitama, Japan;\\
}

\corres{Correspondence: m.shahrbaf46@gmail.com}

\abstract{%
We investigate cold homogeneous matter composed of neutrons, protons, and
$\Lambda$ hyperons within our hyperonic extension of the lowest-order
constrained variational (LOCV) method, hereafter denoted LOCVY.
Our earlier LOCVY calculation, based on two-baryon interactions, is extended
by supplementing the Argonne $v_{18}$ nucleonic interaction with the
Urbana IX three-nucleon force, reduced within the variational framework to
a correlation-weighted density-dependent effective two-nucleon interaction.
The $N\Lambda$ and $\Lambda\Lambda$ interactions are kept unchanged,
allowing the present calculation to isolate the competition between
hyperon-induced softening and nucleonic three-body repulsion.
The energy per baryon is calculated for fixed $\Lambda$ fractions
$Y_{\Lambda}=0$, $0.1$, and $0.2$ in matter with a symmetric nucleonic
component and in the proton-free neutron--$\Lambda$ limit.
Direct differences between calculations with and without the three-body
force quantify its density-dependent contribution, while a complementary
decomposition into $NN$, $N\Lambda$, and $\Lambda\Lambda$ terms identifies
the microscopic origin of the stiffening.
The Urbana contribution becomes increasingly repulsive with density and
opposes, but does not generically remove, the softening associated with a
finite $\Lambda$ content.
We further investigate the saturation properties for several prescribed
$\Lambda$ fractions, with and without the nucleonic three-body force, to
clarify how strangeness and many-body interactions modify the saturation
point and the agreement with empirical nuclear-matter properties.
}
\keyword{hypernuclear matter; equation of state; three-nucleon force;
Urbana IX interaction; lowest-order constrained variational method;
$\Lambda$ hyperons}

\begin{document}
%  \nolinenumbers
%=================================================================
\section{Introduction}
\label{sec:introduction}
%=================================================================

The equation of state (EOS) of dense baryonic matter connects microscopic
strong interactions to the structure and dynamics of compact stars. Although
empirical nuclear information constrains matter near the saturation density
$\rho_0$, substantial uncertainty remains at several times $\rho_0$, where
many-body forces, isospin asymmetry, and additional hadronic degrees of freedom
become increasingly important \cite{LattimerPrakash2016,Oertel2017, Sakuragi:2016jni}. A
microscopic EOS must therefore address both the correlations generated by
realistic baryon--baryon interactions and the many-body contributions that are
not captured by two-nucleon forces alone.

Hyperons provide a particularly important extension of the nuclear sector.
Experimental hypernuclear spectroscopy constrains parts of the hyperon--nucleon
and hyperon--hyperon interactions, but the available information remains far
less complete than for the nucleon--nucleon system
\cite{ALICE:2020mfd, Gal2016, TolosFabbietti2020}. In dense matter, the appearance of hyperons
can lower the Fermi energy of the baryonic system. This typically reduces the
pressure at fixed energy density and softens the EOS
\cite{Zdunik:2003vg, ChatterjeeVidana2016, OertelHyperons2016, Vidana2018, Logoteta2021, Shahrbaf:2019wex}. The
resulting tension between hyperonic degrees of freedom and the existence of
massive neutron stars (NSs) is usually referred to as the hyperon puzzle. Proposed
resolutions include stronger repulsion in hyperonic two-body channels,
hyperon--nucleon--nucleon forces, and changes in the high-density composition
\cite{Lonardoni2015, Maslov:2015msa, Masuda:2015kha, Wirth:2016iwn, Shahrbaf:2019vtf, Shahrbaf:2020uau}.

Three-nucleon forces are independently required in the ordinary nuclear sector.
High-precision two-nucleon interactions provide an accurate description of
nucleon--nucleon scattering data, but calculations based on them alone
generally underbind light nuclei and fail to reproduce the empirical
saturation point of symmetric nuclear matter
\cite{Pieper:2001ap, Akmal1998}. This provides a strong motivation for the
inclusion of three-nucleon interactions in microscopic descriptions of
nuclear matter.
The Urbana IX (UIX) interaction is a well-known nucleonic three-body force (TBF) which combines a long-range two-pion-exchange term
with a phenomenological short-range repulsive contribution
\cite{Fujita:1957zz, Pudliner1997}. Together with the Argonne $v_{18}$ interaction \cite{Wiringa1995},
UIX has been used in variational descriptions of nuclear matter and NSs \cite{Akmal1998}. Within the LOCV framework, an Urbana-type
TBF has been reduced to an effective density-dependent two-body
interaction and applied to nucleonic matter. These studies demonstrated its
importance for saturation, high-density pressure, neutron star (NS) masses, and the
density dependence of the nuclear symmetry energy
\cite{Goudarzi2015,Goudarzi2018}.

The LOCV method is a self-consistent variational approach in which a
state-dependent Jastrow correlation operator is determined by minimizing the
cluster-expanded energy subject to a normalization constraint
\cite{BordbarModarres1998, Modarres:2000nk, Moshfegh:2005ouk, Modarres:2009zz}. In Ref.~\cite{Shahrbaf2019}, the method was extended
to homogeneous matter containing $\Lambda$ hyperons. The calculation employed
the Argonne $v_{18}$ potential for nucleons and phenomenological
spin-parity-dependent $N\Lambda$ and $\Lambda\Lambda$ interactions constrained
by single- and double-$\Lambda$ hypernuclear information
\cite{Rijken1999,Hiyama2006,Hiyama2002,Takahashi2001}. It showed explicitly
that increasing the $\Lambda$ fraction softens the EOS and that the uncertain
odd-state $\Lambda\Lambda$ interaction affects the high-density behavior. In the original development of LOCVY as the hyperonic extension of the
LOCV framework, only two-baryon interactions were included. This choice
was made to isolate as clearly as possible the effects associated with
the introduction of hyperonic degrees of freedom, while also keeping the
initial numerical implementation computationally manageable.

The present work closes this specific gap by incorporating the same
Urbana-type nucleonic three-body treatment used in previous LOCV studies into
the hypernuclear calculation of Ref.~\cite{Shahrbaf2019}. The purpose is not to
repeat the earlier extensive discussion of correlation functions. Instead, we
focus on a controlled comparison of the energy with and without
the nucleonic TBF at fixed composition. We calculate matter with
$Y_{\Lambda}=0$, $0.1$, and $0.2$ for a symmetric nucleonic component and for
the proton-free neutron--$\Lambda$ limit. 

Beyond examining the net modification of the EOS, we analyze the microscopic
origin of these changes in considerably greater detail. In particular, the
interaction energy is decomposed into its $NN$, $N\Lambda$, and
$\Lambda\Lambda$ contributions at representative densities
$\rho_0$, $3\rho_0$, and $5\rho_0$, allowing us to trace how the nucleonic
TBF redistributes the interaction energy among the different
baryonic sectors as the density increases. For the $\Lambda\Lambda$ sector,
we additionally compare the attractive and repulsive prescriptions for the
poorly constrained odd-state interaction, thereby quantifying the sensitivity
of the microscopic energy balance to this remaining uncertainty in the
hyperon--hyperon interaction. We further determine the saturation properties
for several prescribed $\Lambda$ fractions, both with and without the
nucleonic TBF, in order to disentangle the effects of strangeness
and many-body repulsion on the evolution of the saturation point. This analysis
also permits a direct assessment of the recovery of empirical nuclear-matter
saturation properties in the $Y_{\Lambda}=0$ limit.
The resulting EOSs should therefore be regarded as microscopic
fixed-composition benchmarks for isospin-symmetric and asymmetric
hyperonic matter, rather than as complete EOS for
beta-equilibrated stellar matter.

The paper is organized as follows. Section~\ref{sec:framework} summarizes the
LOCVY formalism, the adopted two-baryon interactions, and the density-dependent
effective two-body representation of UIX employed in LOCVY method. The results are presented in
Section~\ref{sec:results}, with emphasis on the energy per baryon, the isolated
three-body contribution, saturation properties, and microscopic
energy decomposition. Section~\ref{sec:conclusions} gives the conclusions and
outlook.

%=================================================================
\section{Theoretical Framework}
\label{sec:framework}
%=================================================================

\subsection{Composition and kinematics}

We consider uniform, zero-temperature matter composed of neutrons, protons,
and $\Lambda$ hyperons. The total baryon density is
\begin{equation}
\rhoB=\rho_n+\rho_p+\rhoL,
\label{eq:rhoB}
\end{equation}
and the particle fractions are
\begin{equation}
x_i=\frac{\rho_i}{\rhoB},
\qquad i\in\{n,p,\Lambda\},
\qquad
x_n+x_p+x_{\Lambda}=1.
\label{eq:fractions}
\end{equation}
We denote the $\Lambda$ fraction by
\begin{equation}
\YN\equiv x_{\Lambda}=\frac{\rhoL}{\rhoB}.
\label{eq:YLambda}
\end{equation}
For the nucleonic subsystem, the proton fraction and asymmetry are
\begin{equation}
x_p^{(N)}=\frac{\rho_p}{\rho_n+\rho_p},
\qquad
\delta_N=\frac{\rho_n-\rho_p}{\rho_n+\rho_p}.
\label{eq:asymmetry}
\end{equation}
The two limiting compositions studied here are
\begin{align}
\delta_N=0:&\quad
\rho_n=\rho_p=\frac{1-\YN}{2}\rhoB,
\label{eq:symmetriccomp}\\
\delta_N=1:&\quad
\rho_p=0,
\qquad
\rho_n=(1-\YN)\rhoB.
\label{eq:neutronlambdacomp}
\end{align}
The second case is referred to below as neutron--$\Lambda$ matter or the
proton-free limit; for $\YN=0$, it reduces to pure neutron matter. The Fermi
momentum of each component is
\begin{equation}
k_{F,i}=(3\pi^2\rho_i)^{1/3}.
\label{eq:kFi}
\end{equation}

\subsection{LOCV energy functional}

The correlated many-body wave function is written in Jastrow form,
\begin{equation}
\Psi(1,\ldots,A)=F(1,\ldots,A)\Phi(1,\ldots,A),
\label{eq:jwave}
\end{equation}
where $\Phi$ is the Slater determinant of the noninteracting multicomponent
Fermi system and
\begin{equation}
F=\mathcal{S}\prod_{i<j} f(ij)
\label{eq:corrop}
\end{equation}
is the symmetrized product of state-dependent two-body correlation operators.
For a channel $\alpha$, the operator is expanded as
\begin{equation}
f(ij)=\sum_{\alpha,p}f_{\alpha}^{(p)}(r_{ij})
O_{\alpha}^{(p)}(ij),
\label{eq:fexpand}
\end{equation}
where $\alpha$ collects the relevant angular-momentum, spin, isospin, and
particle-species labels. The explicit operator structure is given in Ref.~\cite{Shahrbaf2019} and is not
repeated here.

The Hamiltonian used in the present extension is
\begin{equation}
H=\sum_i\frac{p_i^2}{2m_i}
 +\sum_{i<j}V_{ij}^{(2)}
 +\sum_{i<j<k\in N}V_{ijk}^{\mathrm{UIX}},
\label{eq:Hamiltonian}
\end{equation}
where $V_{ij}^{(2)}$ contains the $NN$, $N\Lambda$, and
$\Lambda\Lambda$ two-body interactions. The last sum acts only on triples of
nucleons. Thus, no $NN\Lambda$, $N\Lambda\Lambda$, or
$\Lambda\Lambda\Lambda$ hyperonic three-baryon forces are included
in the present calculation, as the investigation of hyperonic
three-body interactions lies beyond the scope of this study.

At the lowest cluster order, the energy per baryon is
\begin{equation}
\frac{E}{A}=E_1+E_2,
\label{eq:Ecluster}
\end{equation}
with the one-body term
\begin{equation}
E_1=\sum_{i=n,p,\Lambda}x_i
\frac{3\hbar^2 k_{F,i}^2}{10m_i},
\label{eq:E1}
\end{equation}
and the correlated two-body term
\begin{equation}
E_2=\frac{1}{2A}\sum_{ij}
\avg{ij\left|W(12)\right|ij-ji},
\label{eq:E2}
\end{equation}
where
\begin{equation}
W(12)=-\frac{\hbar^2}{2\mu_{12}}
\left[f(12),\left[\nabla_{12}^2,f(12)\right]\right]
+f(12)V_{12}^{\mathrm{eff}}f(12).
\label{eq:W12}
\end{equation}
Here $\mu_{12}$ is the reduced mass appropriate to the pair. In the nucleonic
sector, $V_{12}^{\mathrm{eff}}$ includes the density-dependent contribution
generated by UIX; in the hyperonic sectors it reduces to the adopted two-body
interaction.

Variation of $E_2$ is performed subject to the LOCV normalization constraint,
which can be written schematically as
\begin{equation}
\frac{1}{A}\sum_{ij}
\avg{ij\left|F_{P}^{2}(12)-f^{2}(12)\right|ij-ji}=0.
\label{eq:constraint}
\end{equation}
The Pauli function $F_P$ is applied to indistinguishable fermion pairs, while
distinguishable pairs are not subjected to an exchange restriction. The
variational equations are solved independently at each density and composition.

For the discussion of the microscopic mechanism, the energy per baryon
is decomposed as
\begin{equation}
\frac{E}{A}
=
E_{\rm kin}
+
E_{NN}
+
E_{N\Lambda}
+
E_{\Lambda\Lambda},
\label{eq:energy_decomposition}
\end{equation}
where $E_{\rm kin}$ denotes the total one-body kinetic-energy contribution,
while $E_{NN}$, $E_{N\Lambda}$, and $E_{\Lambda\Lambda}$ denote the
correlated interaction-energy contributions associated with the corresponding
baryon pairs. 
%A further partial-wave decomposition is introduced through \begin{equation} E_{BB'}=\sum_{\alpha}E_{\alpha}^{BB'}, \qquad B,B'\in\{N,\Lambda\}. \label{eq:partialdecomp} \end{equation}

\subsection{Two-baryon interactions}

The nucleonic two-body interaction is Argonne $v_{18}$
\cite{Wiringa1995}. 
The $N\Lambda$ interaction is described by the central, spin- and
parity-dependent three-range Gaussian potential adopted in
Ref.~\cite{Shahrbaf2019},
\begin{equation}
V_{N\Lambda}^{(\alpha)}(r)
=
\sum_{i=1}^{3}
v_i^{(\alpha)}
\exp\left[
-\left(\frac{r}{\beta_i}\right)^2
\right],
\label{eq:NLpotential}
\end{equation}
where $\alpha\in\{{}^{1}E,{}^{3}E,{}^{1}O,{}^{3}O\}$ denotes the
spin-parity channel. The superscripts $1$ and $3$ correspond,
respectively, to the spin-singlet ($S=0$) and spin-triplet ($S=1$)
states, according to the spin multiplicity $2S+1$, while $E$ and $O$
denote even- and odd-parity states, corresponding to even and odd
values of the relative orbital angular momentum $L$, respectively.
The parameters $v_i^{(\alpha)}$ specify the interaction strength in
each spin-parity channel, whereas $\beta_i$ determine the ranges of
the three Gaussian components. Its parameters were
constructed to reproduce selected single-$\Lambda$ hypernuclear observables
and the main features of the Nijmegen soft-core interaction
\cite{Rijken1999,Hiyama2006}.

The even- and odd-parity $\Lambda\Lambda$ interactions are written as
\begin{equation}
V_{\Lambda\Lambda}^{(\pi)}(r)
=
\sum_{i=1}^{3}
\left[
v_i^{(\pi)}
+
v_{\sigma,i}^{(\pi)}
\boldsymbol{\sigma}_1\!\cdot\!\boldsymbol{\sigma}_2
\right]
\exp\left(-\mu_i^{(\pi)}r^2\right),
\label{eq:LLpotential}
\end{equation}
where $\pi=E,O$ denotes the even- and odd-parity sectors,
respectively. The coefficients $v_i^{(\pi)}$ and
$v_{\sigma,i}^{(\pi)}$ specify, respectively, the central and
spin-dependent strengths of the $i$th Gaussian component in a given
parity channel, while $\mu_i^{(\pi)}$ determines the corresponding
inverse squared range. The operators
$\boldsymbol{\sigma}_1$ and $\boldsymbol{\sigma}_2$ are the Pauli spin
operators of the two $\Lambda$ hyperons. The even-state interaction is constrained by double-$\Lambda$
hypernuclear information, including the NAGARA event
\cite{Hiyama2002,Takahashi2001}. The odd-state interaction remains poorly constrained. We therefore
retain both the attractive and repulsive prescriptions introduced in
Ref.~\cite{Shahrbaf2019} in order to assess the sensitivity of the
EOS and, in particular, the $\Lambda\Lambda$ contribution to this
remaining interaction uncertainty.

%=================================================================
\subsection{Urbana IX Three-Nucleon Force}
\label{subsec:urbana}
%=================================================================

The UIX three-nucleon interaction consists of a
Fujita--Miyazawa-type two-pion-exchange term
\cite{Fujita:1957zz} and a phenomenological short-range repulsive
contribution \cite{Pudliner1995, Pudliner1997},
\begin{equation}
V_{123}^{\mathrm{UIX}}
=V_{123}^{2\pi}+V_{123}^{R}.
\label{eq:UIXsum}
\end{equation}
 
The details about two-pion-exchange and the short-range parts are given in the
previous LOCV implementation \cite{Goudarzi2015,Goudarzi2018}. 

Following the established LOCV prescription, the third nucleon is averaged out
with the appropriate exchange and correlation operators. Schematically,
\begin{equation}
\VTBF(12;\rho_n,\rho_p)
=\operatorname{Tr}_{3}\int\dd^3r_3\,
\mathcal{C}_{123}\,V_{123}^{\mathrm{UIX}},
\label{eq:V3eff}
\end{equation}
where $\mathcal{C}_{123}$ represents the correlation dressing, statistical
weights, and exchange structure used in the averaging procedure. The effective
nucleonic interaction entering Eq.~\eqref{eq:W12} is then
\begin{equation}
V_{NN}^{\mathrm{eff}}(12;\rho_n,\rho_p)
=V_{NN}^{\mathrm{AV18}}(12)+\VTBF(12;\rho_n,\rho_p).
\label{eq:VNNtotal}
\end{equation}
The full spin-isospin traces and radial expressions are those of
Refs.~\cite{Goudarzi2015,Goudarzi2018}. 

Two points are essential. First, the UIX correction depends on the neutron and
proton densities, not directly on the total baryon density. At fixed $\rhoB$,
increasing $\YN$ reduces the nucleon density,
\begin{equation}
\rho_N=\rho_n+\rho_p=(1-\YN)\rhoB,
\label{eq:rhoN}
\end{equation}
and therefore changes the direct strength of the nucleonic three-body
contribution. Second, the effective interaction is inserted before the
variational solution is obtained. Consequently, switching on UIX can also
modify the optimized two-body contributions indirectly; therefore, it does not act as a
simple additive shift.

%=================================================================
\section{Results and Discussion}
\label{sec:results}
%=================================================================

The calculations are performed for $\YN=0$, $0.1$, and $0.2$ in the two
compositions defined by Eqs.~\eqref{eq:symmetriccomp} and
\eqref{eq:neutronlambdacomp}. For every composition, the same density grid and
hyperonic interactions are used in the two-body-only and AV18+UIX calculations.
This one-to-one setup allows differences to be attributed to the nucleonic
three-body sector without contamination from changes in the remaining model
inputs. 

\subsection{Energy per baryon: Direct comparison with the two-body baseline}
\label{subsec:energy}

\begin{figure}[tbp]
    \centering

    \includegraphics[width=0.48\textwidth]{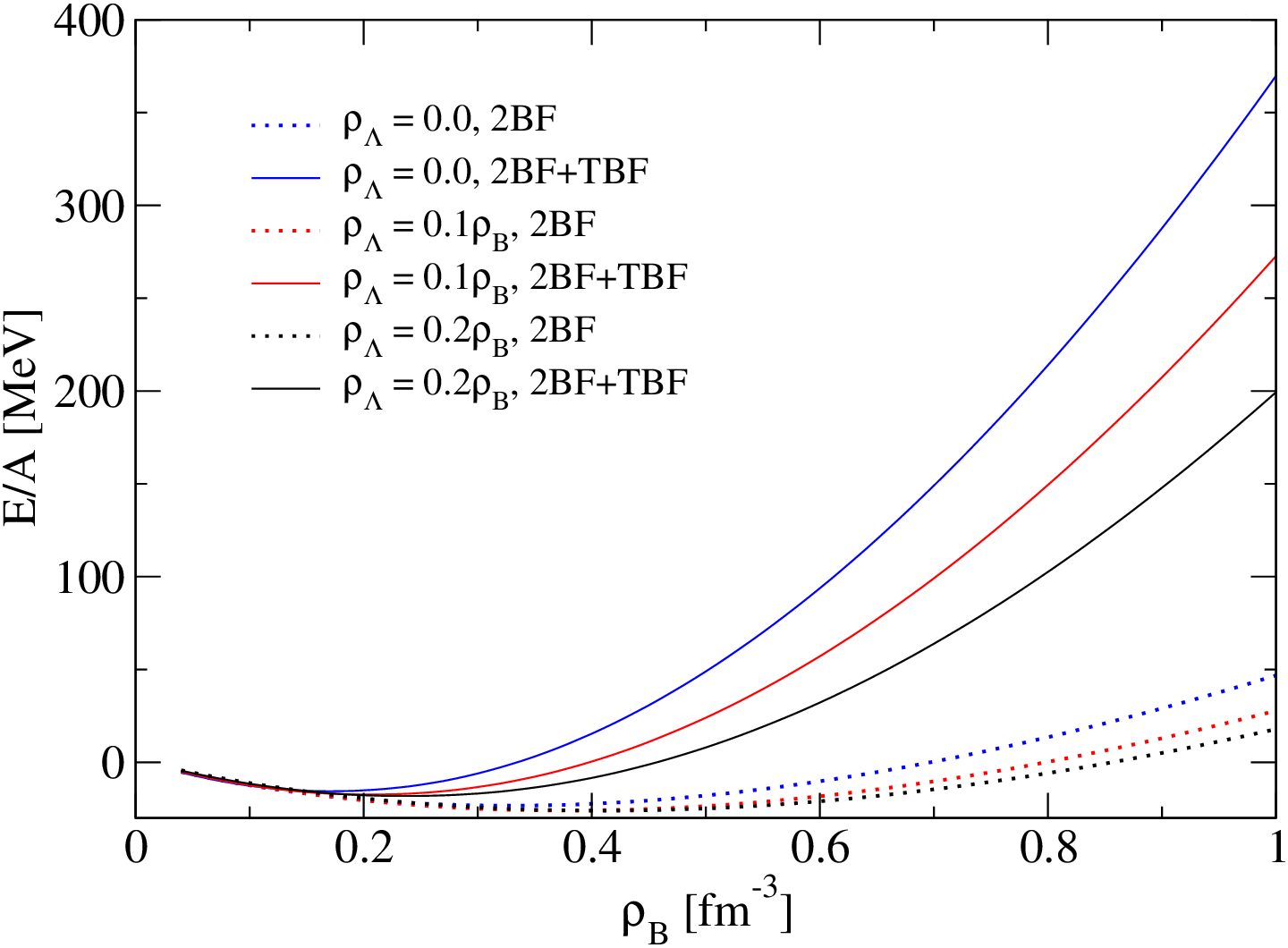}
    \hfill
    \includegraphics[width=0.48\textwidth]{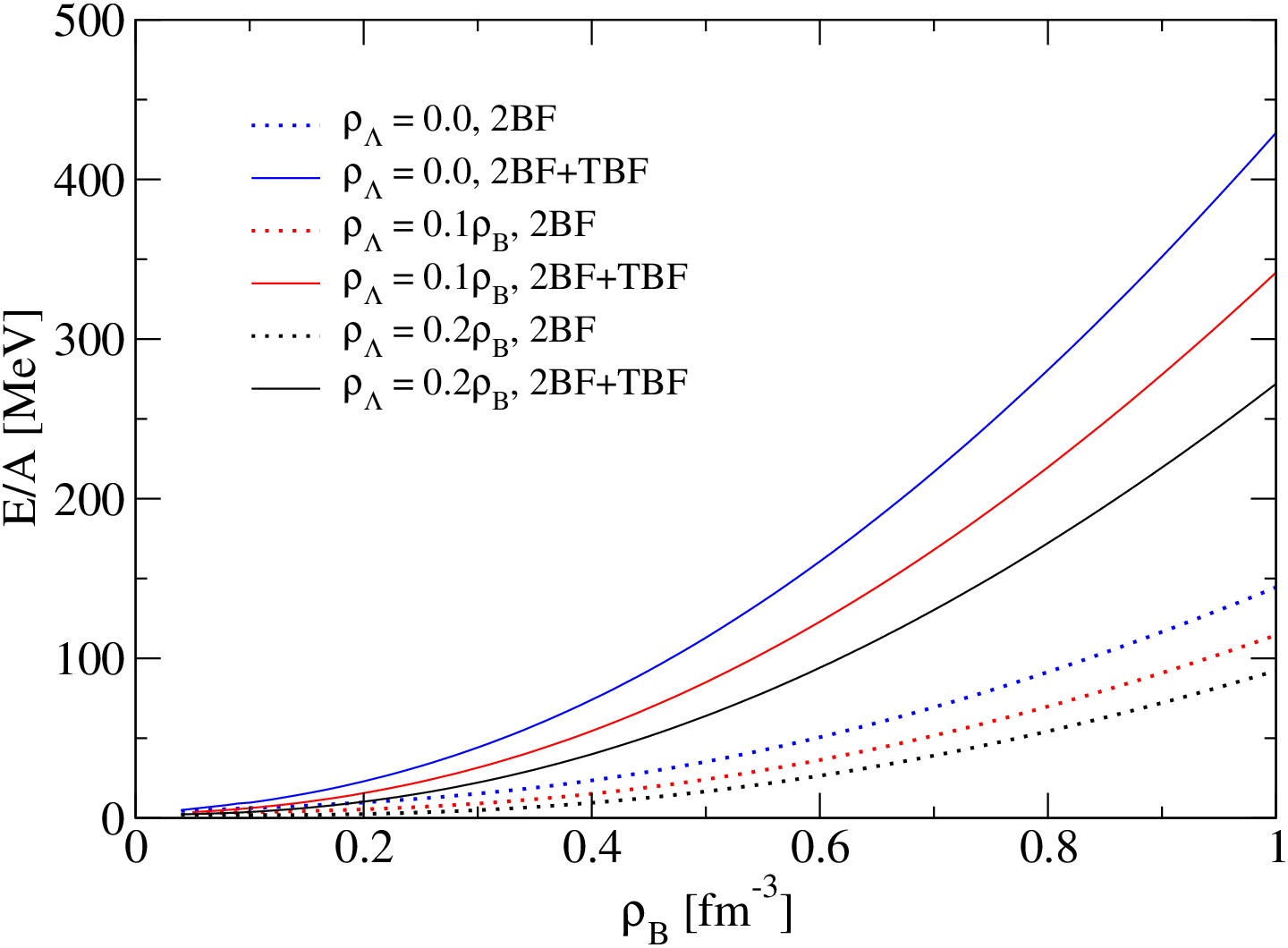}

    \vspace{0.3cm}

    \includegraphics[width=0.48\textwidth]{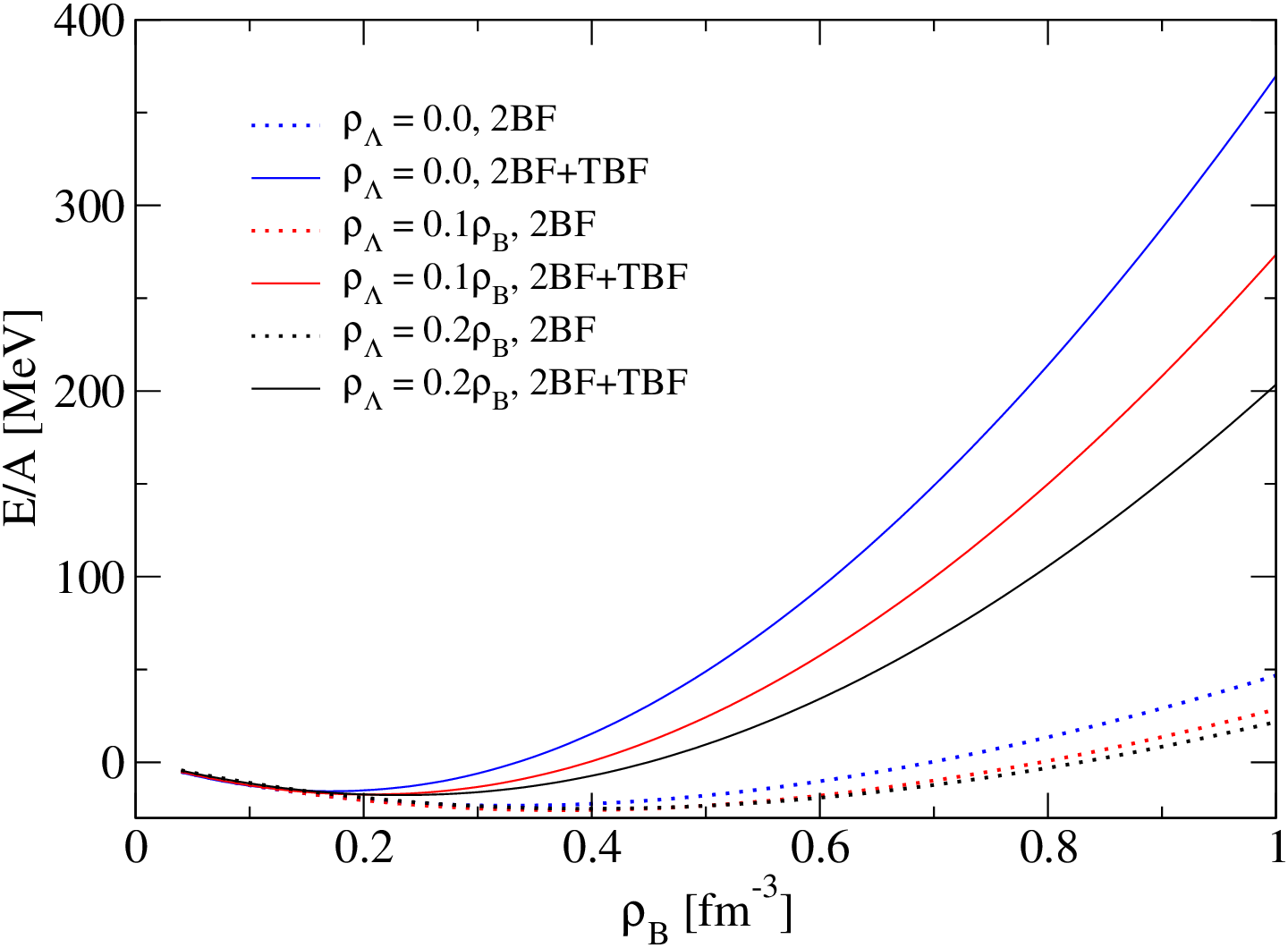}
    \hfill
    \includegraphics[width=0.48\textwidth]{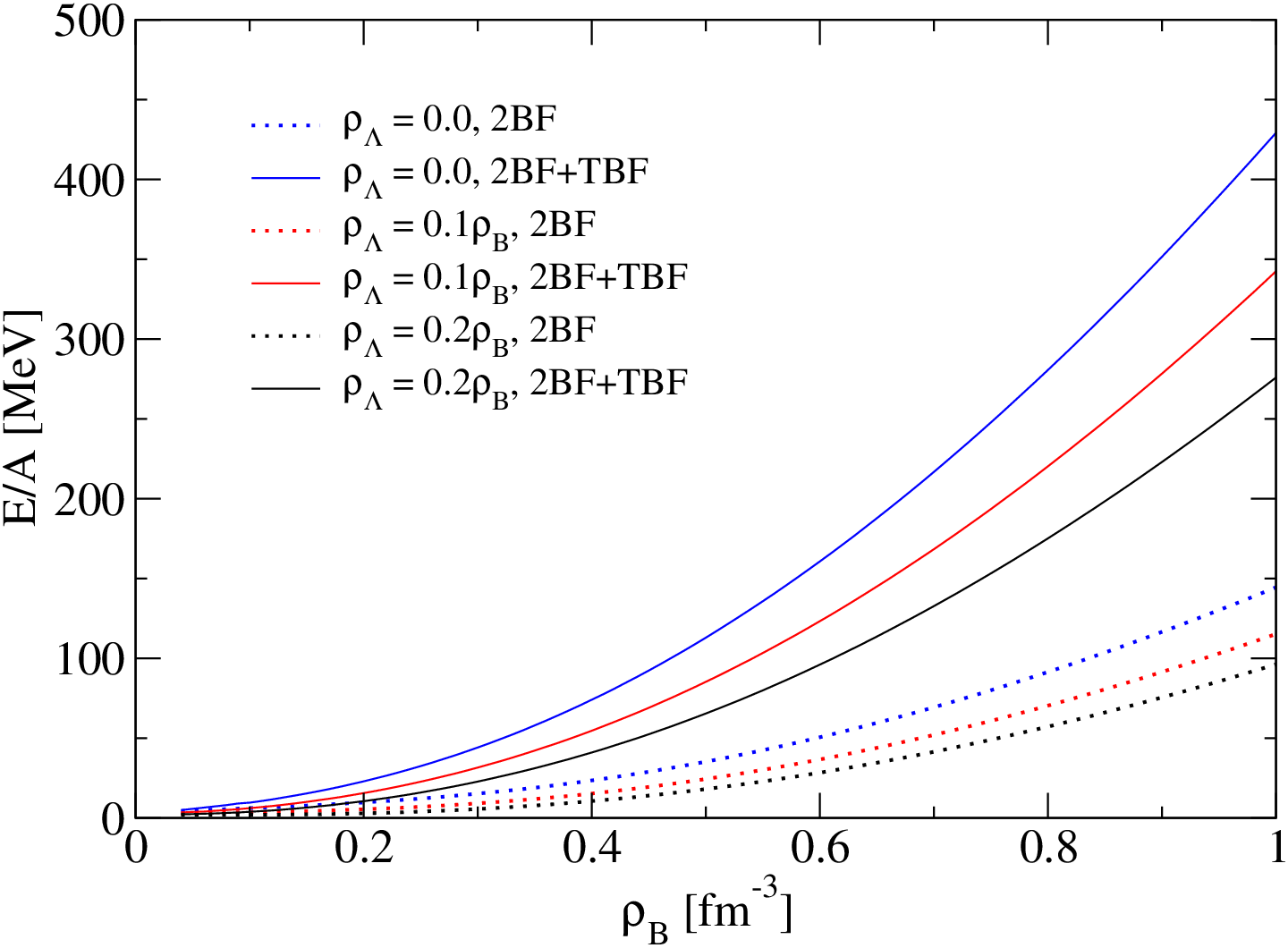}

    \caption{
    Energy per baryon as a function of baryon density for
$Y_{\Lambda}=0$, $0.1$, and $0.2$. Left panels show matter with a symmetric
nucleonic component while right panels show proton-free neutron--$\Lambda$ matter. Upper panels correspond to the attractive interaction for the odd-state part of $\Lambda\Lambda$ interaction and lower panels correspond to the repulsive interaction for the odd-state part of $\Lambda\Lambda$ interaction. The rest-mass
contribution is not included.
    }
    \label{fig:four_panel}
\end{figure}

Figure~\ref{fig:four_panel} illustrates the competing effects of
$\Lambda$ hyperons and the UIX TBF on the energy per
baryon. At fixed baryon density, increasing the $\Lambda$ fraction from
$Y_{\Lambda}=0$ to $0.2$ systematically lowers $E/A$, demonstrating the
softening produced by the replacement of nucleons with $\Lambda$
hyperons. In contrast, the inclusion of UIX generates an increasingly
repulsive contribution with density, leading to a pronounced stiffening
of the EOS at supra-saturation densities. The proton-free
neutron--$\Lambda$ system is less bound than matter with a symmetric
nucleonic component, reflecting the additional energy associated with
nucleonic isospin asymmetry. The attractive and repulsive prescriptions
for the odd-state part of the $\Lambda\Lambda$ interaction preserve the
same overall hierarchy of the curves, indicating that, for the
compositions considered here, their effect is smaller than the
competition between hyperonic softening and the density-dependent
repulsion generated by UIX.

 We define
\begin{equation}
\Delta E_{\mathrm{TBF}}(\rhoB,\YN,\delta_N)
=\left(\frac{E}{A}\right)_{\mathrm{2BF+TBF}}
-\left(\frac{E}{A}\right)_{\mathrm{2BF}},
\label{eq:DeltaETBF}
\end{equation}
this quantity isolates the net effect of UIX after reoptimization of the
LOCV solution.

\begin{figure}[tbp]
    \centering

    \includegraphics[width=0.48\textwidth]{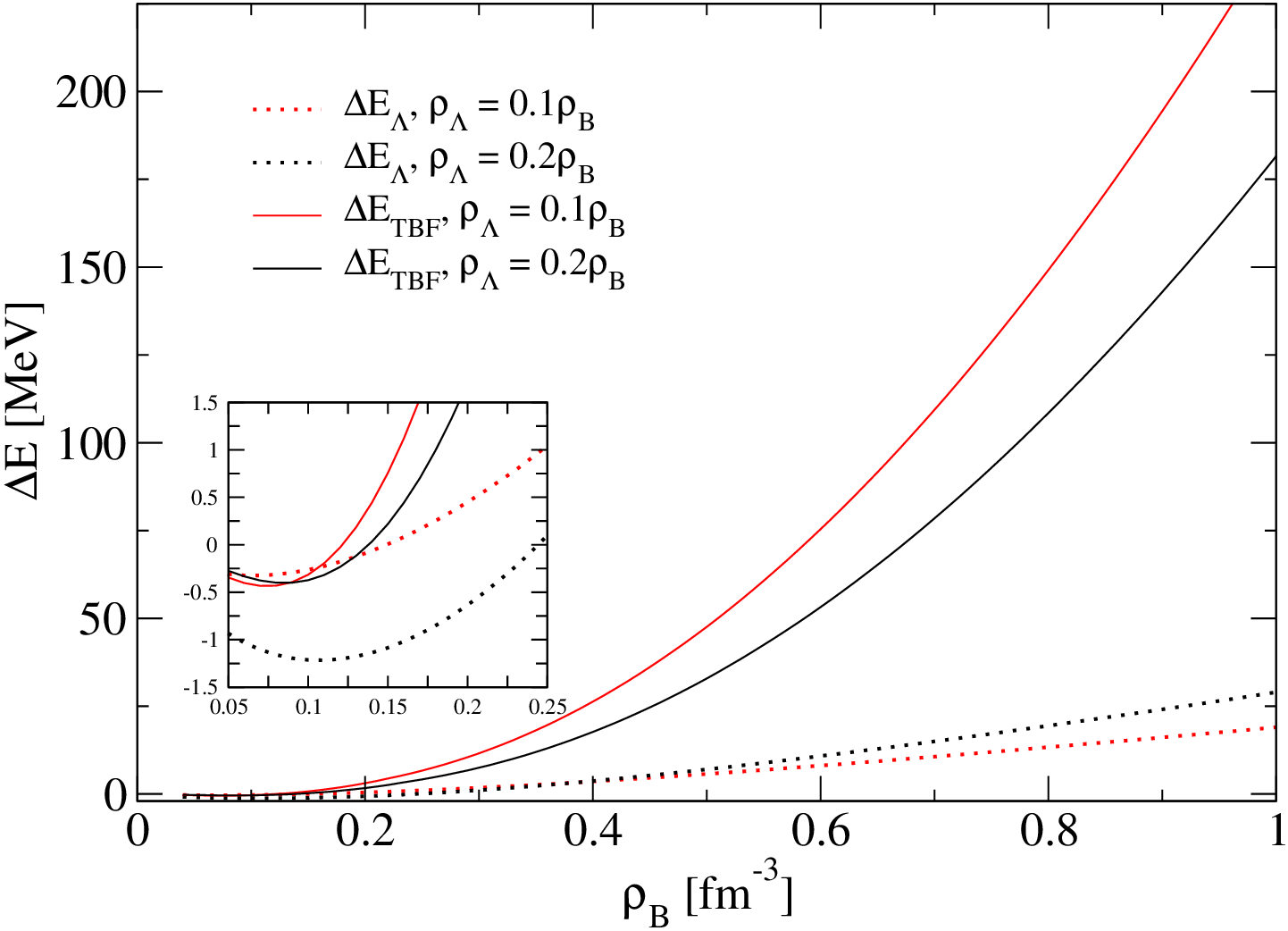}
    \hfill
    \includegraphics[width=0.48\textwidth]{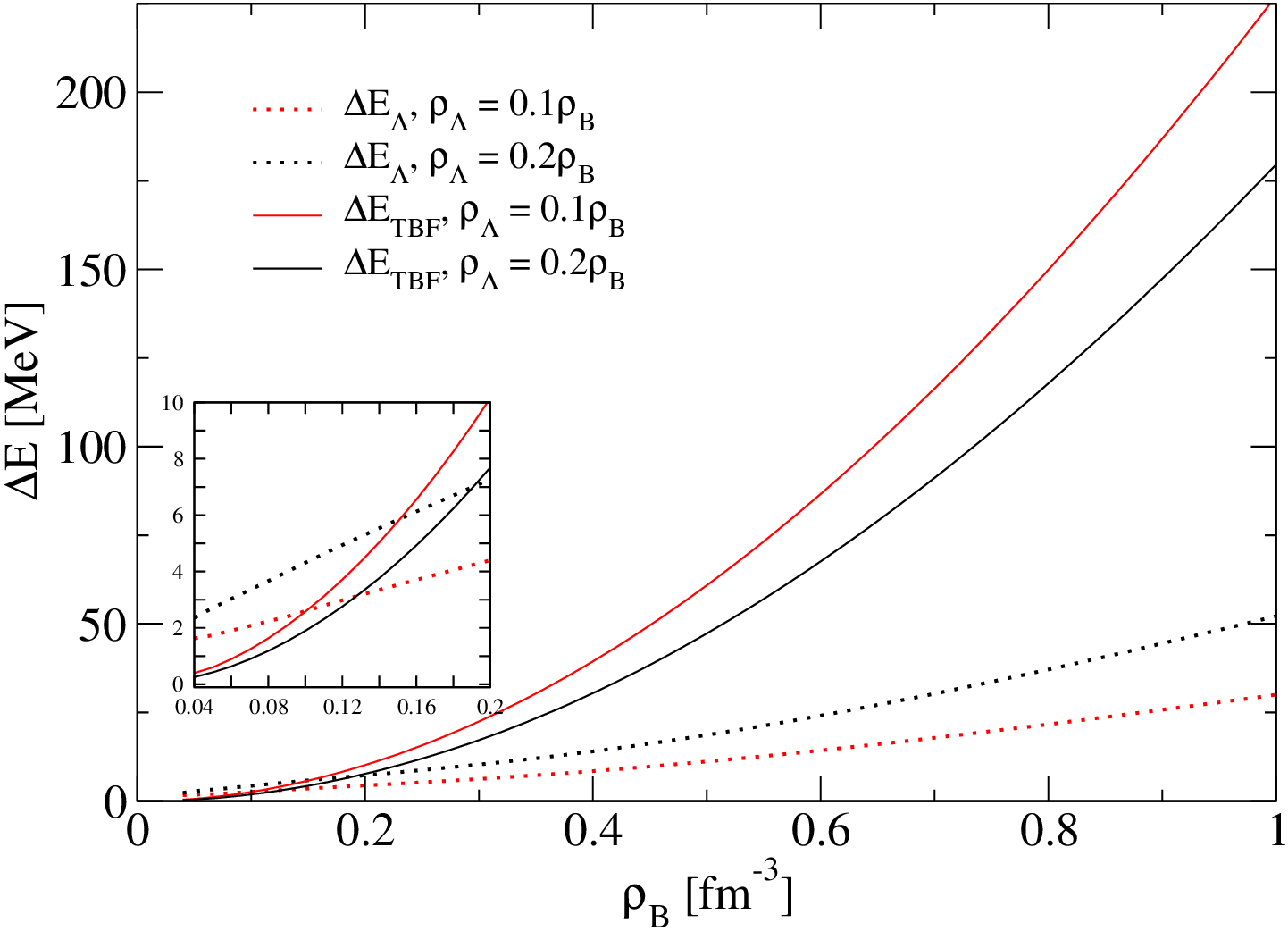}

    \caption{Density dependence of the nucleonic three-body-force
contribution $\Delta E_{\mathrm{TBF}}$, defined in
Eq.~\eqref{eq:DeltaETBF}, shown by solid lines, together with the
hyperonic energy shift $\Delta E_{\Lambda}$, defined in
Eq.~\eqref{eq:delta_lambda}, shown by dotted lines. The left panel
corresponds to matter with a symmetric nucleonic component, while
the right panel shows proton-free neutron--$\Lambda$ matter.
The insets magnify the low-density region.}
    \label{fig:deltas}
\end{figure}

Figure~\ref{fig:deltas} compares the isolated contribution of the
Urbana-type nucleonic TBF, $\Delta E_{\mathrm{TBF}}$, with
the change in the energy per baryon produced by introducing a prescribed
$\Lambda$ fraction,
\begin{equation}
\Delta E_{\Lambda}
\left(\rho_B,Y_{\Lambda},\delta_N\right)
=
\frac{E}{A}\left(\rho_B,Y_{\Lambda}=0,\delta_N\right)
-
\frac{E}{A}\left(\rho_B,Y_{\Lambda},\delta_N\right).
\label{eq:delta_lambda}
\end{equation}
With this convention, $\Delta E_{\Lambda}>0$ means that the
rest-mass-subtracted energy per baryon of the prescribed hyperonic
composition is lower than that of the corresponding purely nucleonic
reference, whereas $\Delta E_{\Lambda}<0$ indicates the opposite.
Since the particle rest masses are not included and the compositions
are imposed rather than determined by chemical equilibrium,
$\Delta E_{\Lambda}$ should not be interpreted as a criterion for
the thermodynamic stability or physical onset of $\Lambda$ hyperons.

For matter with a symmetric nucleonic component,
$\Delta E_{\Lambda}$ changes from negative to positive at a
zero-crossing density defined by
\begin{equation}
\Delta E_{\Lambda}
\left(\rho_{\mathrm{zc}},Y_{\Lambda},\delta_N\right)=0.
\label{eq:zero_crossing}
\end{equation}
 
 In the two-body calculation, the sign change occurs within
$0.14<\rho_{\mathrm{zc}}<0.15~\mathrm{fm}^{-3}$ for
$Y_{\Lambda}=0.1$ and within
$0.24<\rho_{\mathrm{zc}}<0.25~\mathrm{fm}^{-3}$ for
$Y_{\Lambda}=0.2$. After including UIX, the corresponding intervals
decrease to $0.11<\rho_{\mathrm{zc}}<0.12~\mathrm{fm}^{-3}$ and
$0.14<\rho_{\mathrm{zc}}<0.15~\mathrm{fm}^{-3}$, respectively. Thus, the inclusion of the nucleonic TBF shifts the zero crossing of
the rest-mass-subtracted energy difference to a lower total baryon
density.
This occurs because UIX acts only among nucleons and therefore raises
the energy of the purely nucleonic reference system more strongly than
that of hyperonic matter, where the nucleon density is reduced. The
larger value of $\rho_{\mathrm{zc}}$ for $Y_{\Lambda}=0.2$ also shows
that the low-density energetic cost of imposing a larger hyperon
fraction persists to higher density.

No zero crossing is found for proton-free neutron--$\Lambda$ matter
within the density range considered. In this case,
$\Delta E_{\Lambda}$ remains positive, indicating that replacing part
of the neutron component by $\Lambda$ hyperons lowers the energy
throughout the calculated interval. The contrast with symmetric matter
demonstrates that the hyperon-induced energy change depends strongly on
the isospin composition of the nucleonic background. These
zero-crossing densities refer only to comparisons between fixed
compositions and must not be identified with the physical onset density
of $\Lambda$ hyperons in beta-equilibrated matter.

The solid curves further show that $\Delta E_{\mathrm{TBF}}$ grows
rapidly with density and becomes much larger than
$\lvert\Delta E_{\Lambda}\rvert$ in the supra-saturation region for both
values of $Y_{\Lambda}$ and for both nucleonic compositions. The smaller
three-body contribution obtained for $Y_{\Lambda}=0.2$ than for
$Y_{\Lambda}=0.1$ is qualitatively consistent with
$\rho_N=(1-Y_{\Lambda})\rho_B$, since increasing the hyperon fraction
reduces the density of nucleons on which UIX acts. The reduction is not
a simple linear scaling with $1-Y_{\Lambda}$, however, because the
three-body contribution is nonlinear in density and the LOCVY
correlations readjust self-consistently with composition. The similar
density dependence observed in symmetric and proton-free matter
indicates that the total nucleon density is the dominant control
parameter, while the nucleonic isospin asymmetry has a smaller effect.

Overall, the results demonstrate that the Urbana-type nucleonic TBF is
essential for correcting the excessive binding and high saturation density
generated by the two-body calculation. However, because this repulsion acts
only among nucleons, its relative importance decreases as the $\Lambda$
fraction grows. The persistence of composition-dependent shifts toward
higher equilibrium densities and stronger binding at finite $Y_\Lambda$
shows that the nucleonic TBF does not eliminate the changes induced by
strangeness in the saturation properties. Determining whether additional
repulsion from genuine hyperonic three-body interactions is required in
stellar matter will require their consistent implementation in
$\beta$-equilibrated matter and subsequent NS structure calculations. Previous
studies indicate that additional repulsion in the hyperonic sector may be
required to reconcile hyperonic matter with the observed high NS
masses \cite{Vidana:2022tlx, Tong:2025sui, Gerstung:2020ktv}. This motivates a future extension of the LOCVY framework to include
$NN\Lambda$, $N\Lambda\Lambda$, and $\Lambda\Lambda\Lambda$ three-baryon
forces and to assess their effects consistently in beta-stable stellar
matter.

\subsection{Saturation properties and symmetry-energy coefficients}
\label{sec:saturation}

To quantify the influence of the nucleonic TBF beyond the direct comparison of the energy curves, we extract the characteristic bulk parameters of hypernuclear matter at fixed $\Lambda$ fraction. 

For each value of $Y_{\Lambda}$, the saturation density is defined by the minimum of the energy per baryon of the symmetric nucleonic component,
\begin{equation}
\left.
\frac{\partial E_0(\rho_B,Y_{\Lambda})}
{\partial \rho_B}
\right|_{\rho_B=\rho_{\mathrm{sat}}}=0,
\label{eq:saturation_condition}
\end{equation}
where
\begin{equation}
E_0(\rho_B,Y_{\Lambda})
\equiv
E(\rho_B,\delta_N=0,Y_{\Lambda}).
\end{equation}
The corresponding saturation energy is
\begin{equation}
E_{\mathrm{sat}}(Y_{\Lambda})
=
E_0\!\left(\rho_{\mathrm{sat}},Y_{\Lambda}\right).
\end{equation}

The curvature and skewness of the symmetric-matter energy per particle around its minimum, characterized by the incompressibility and skewness parameters, respectively, are given by
\begin{equation}
K_{\mathrm{sat}}(Y_{\Lambda})
=
9\rho_{\mathrm{sat}}^2
\left.
\frac{\partial^2 E_0}
{\partial\rho_B^2}
\right|_{\rho_{\mathrm{sat}}},
\label{eq:Ksat}
\end{equation}
and
\begin{equation}
Q_{\mathrm{sat}}(Y_{\Lambda})
=
27\rho_{\mathrm{sat}}^3
\left.
\frac{\partial^3 E_0}
{\partial\rho_B^3}
\right|_{\rho_{\mathrm{sat}}},
\label{eq:Qsat}
\end{equation}
respectively. Equivalently, introducing
\begin{equation}
\chi=
\frac{\rho_B-\rho_{\mathrm{sat}}}
{3\rho_{\mathrm{sat}}},
\end{equation}
the local density expansion may be written as
\begin{equation}
\begin{aligned}
E_0(\rho_B,Y_{\Lambda})
={}&
E_{\mathrm{sat}}(Y_{\Lambda})
+\frac{1}{2}K_{\mathrm{sat}}(Y_{\Lambda})\chi^2
\\
&+
\frac{1}{6}Q_{\mathrm{sat}}(Y_{\Lambda})\chi^3
+\mathcal{O}(\chi^4).
\end{aligned}
\label{eq:E0_expansion}
\end{equation}
The linear term is absent because the expansion is performed around the equilibrium density defined by Eq.~\eqref{eq:saturation_condition}.

The dependence of the energy per baryon on the nucleonic isospin
asymmetry is commonly characterized through the nuclear symmetry
energy. For ordinary nucleonic matter, the quadratic symmetry-energy
coefficient is defined as
\begin{equation}
S_2(\rho_B)
=
\frac{1}{2}
\left.
\frac{\partial^2 E(\rho_B,\delta_N)}
{\partial\delta_N^2}
\right|_{\delta_N=0}.
\label{eq:common_symmetry_definition}
\end{equation}
Within the frequently used parabolic approximation, higher-order
terms in $\delta_N$ are neglected, and the symmetry energy is commonly
estimated as
\begin{equation}
S_2(\rho_B)
\simeq
E(\rho_B,\delta_N=1)
-
E(\rho_B,\delta_N=0).
\label{eq:parabolic_symmetry_energy}
\end{equation}

For hyperonic matter, we generalize this definition by considering
variations of the nucleonic isospin asymmetry at fixed total baryon
density $\rho_B$ and fixed $\Lambda$ fraction $Y_{\Lambda}$. The energy
per baryon can then be expanded as
\begin{equation}
\begin{aligned}
E(\rho_B,\delta_N,Y_{\Lambda})
={}&
E_0(\rho_B,Y_{\Lambda})
+
S_2(\rho_B,Y_{\Lambda})\delta_N^2
\\
&+
S_4(\rho_B,Y_{\Lambda})\delta_N^4
+
\mathcal{O}(\delta_N^6),
\end{aligned}
\label{eq:isospin_expansion}
\end{equation}
where $S_2(\rho_B,Y_{\Lambda})$ represents the generalized quadratic
symmetry-energy coefficient at fixed $Y_{\Lambda}$, while
$S_4(\rho_B,Y_{\Lambda})$ quantifies the leading correction to the
parabolic approximation. Odd powers of $\delta_N$ are absent when
charge-symmetry-breaking effects are neglected.

In the present calculation, the energy is evaluated at
$\delta_N=0$, $0.5$, and $1$, corresponding to nucleonic proton fractions
$x_p^{(N)}=0.5$, $0.25$, and $0$, respectively. We define
\begin{equation}
D_{1/2}
=
E(\rho_B,\delta_N=0.5,Y_{\Lambda})
-
E(\rho_B,\delta_N=0,Y_{\Lambda}),
\end{equation}
and
\begin{equation}
D_1
=
E(\rho_B,\delta_N=1,Y_{\Lambda})
-
E(\rho_B,\delta_N=0,Y_{\Lambda}).
\end{equation}
Neglecting terms of order $\mathcal{O}(\delta_N^6)$ and higher,
the expansion in Eq.~\eqref{eq:isospin_expansion} gives
\begin{equation}
D_{1/2}
=
\frac{1}{4}S_2
+
\frac{1}{16}S_4,
\end{equation}
and
\begin{equation}
D_1
=
S_2+S_4.
\end{equation}
Solving these two equations yields
\begin{equation}
S_2(\rho_B,Y_{\Lambda})
=
\frac{16D_{1/2}-D_1}{3},
\label{eq:S2_three_asymmetries}
\end{equation}
and
\begin{equation}
S_4(\rho_B,Y_{\Lambda})
=
\frac{4D_1-16D_{1/2}}{3}.
\label{eq:S4_three_asymmetries}
\end{equation}
This procedure allows the quadratic symmetry energy to be extracted without assuming in advance that the parabolic approximation is exact. The magnitude of $S_4$ provides an estimate of the leading quartic
correction to the parabolic approximation.

The symmetry-energy parameters at the saturation density of each fixed-$Y_{\Lambda}$ system are defined by
\begin{equation}
J(Y_{\Lambda})
=
S_2\!\left(\rho_{\mathrm{sat}},Y_{\Lambda}\right),
\label{eq:J_definition}
\end{equation}
\begin{equation}
L(Y_{\Lambda})
=
3\rho_{\mathrm{sat}}
\left.
\frac{\partial S_2}
{\partial\rho_B}
\right|_{\rho_{\mathrm{sat}}},
\label{eq:L_definition}
\end{equation}
and
\begin{equation}
K_{\mathrm{sym}}(Y_{\Lambda})
=
9\rho_{\mathrm{sat}}^2
\left.
\frac{\partial^2 S_2}
{\partial\rho_B^2}
\right|_{\rho_{\mathrm{sat}}}.
\label{eq:Ksym_definition}
\end{equation}
Accordingly, the symmetry energy can be expanded locally as
\begin{equation}
\begin{aligned}
S_2(\rho_B,Y_{\Lambda})
={}&
J(Y_{\Lambda})
+L(Y_{\Lambda})\chi
\\
&+
\frac{1}{2}K_{\mathrm{sym}}(Y_{\Lambda})\chi^2
+\mathcal{O}(\chi^3).
\end{aligned}
\label{eq:S2_expansion}
\end{equation}

For $Y_{\Lambda}>0$, the quantities introduced above should be understood as composition-dependent generalized bulk coefficients of hypernuclear matter at fixed $\Lambda$ fraction. Only their values at $Y_{\Lambda}=0$ correspond directly to the conventional saturation and symmetry-energy parameters of purely nucleonic matter.

Numerically, the coefficients are extracted from local polynomial fits
to the calculated points surrounding the minimum of each
symmetric-matter curve. The fitting interval is selected so as to
remain within the locally smooth region of the corresponding EOS,
and the stability of the extracted coefficients is examined by
varying the polynomial order and fitting window. The symmetry-energy
coefficient $S_2$ is first determined at each available density from
Eq.~\eqref{eq:S2_three_asymmetries} and is then fitted locally to
extract $J$, $L$, and $K_{\mathrm{sym}}$. Higher-derivative quantities,
particularly $Q_{\mathrm{sat}}$ and $K_{\mathrm{sym}}$, are consequently
more sensitive to the fitting prescription.

\begin{table}[H]
\caption{Saturation and generalized symmetry-energy coefficients for different fixed $\Lambda$ fractions. Densities are given in $\mathrm{fm}^{-3}$ and all remaining quantities in MeV.}
\label{tab:saturation}
\centering
\begin{tabular}{ccccccccc}
\toprule
$Y_\Lambda$ &
Interaction &
$\rho_\mathrm{sat}$ &
$E_\mathrm{sat}$ &
$K_\mathrm{sat}$ &
$Q_\mathrm{sat}$ &
$J$ &
$L$ &
$K_\mathrm{sym}$ \\
\midrule
0.0 & 2BF &
0.3273 & -23.3729 &
370.9 & -72 &
38.01 & 74.4 & -40 \\
0.0 & 2BF+UIX &
0.1710 & -15.6434 &
304.6 & -233 &
31.17 & 79.1 & -57 \\
0.1 & 2BF &
0.3775 & -26.0172 &
433.8 & -324 &
37.04 & 76.5 & -66 \\
0.1 & 2BF+UIX &
0.2046 & -17.4584 &
333.5 & -247 &
31.34 & 81.0 & -61 \\
0.2 & 2BF &
0.4093 & -26.0877 &
444.9 & -324 &
33.85 & 71.9 & -56 \\
0.2 & 2BF+UIX$^\ast$ &
0.2402 & -18.1574 &
380.3 & -471 &
30.79 & 83.0 & -71 \\
\bottomrule
\end{tabular}

\noindent $^\ast$For $Y_\Lambda=0.2$ with 2BF+UIX, the isoscalar fit was
restricted to the smooth interval
$0.230\leq\rho_B\leq0.250~\mathrm{fm}^{-3}$.
The quoted $Q_\mathrm{sat}$ is particularly fit-sensitive, with an
estimated local-fit spread of approximately $65~\mathrm{MeV}$.
\end{table}

Table~\ref{tab:saturation} summarizes the saturation properties and the
generalized symmetry-energy coefficients obtained at fixed
$\Lambda$ fractions, both with the AV18 two-body interaction alone and
after inclusion of the Urbana-type nucleonic TBF. The
results reveal a systematic competition between the additional
repulsion generated by the nucleonic three-body interaction and the
softening associated with the replacement of nucleons by $\Lambda$
hyperons. For purely nucleonic symmetric matter, corresponding to
$Y_{\Lambda}=0$, the two-body calculation produces a deeply bound
equilibrium state at a considerably high density,
\begin{equation}
\rho_{\mathrm{sat}}=0.3273~\mathrm{fm}^{-3},
\qquad
E_{\mathrm{sat}}=-23.3729~\mathrm{MeV}.
\end{equation}
These values reflect the excessive attraction of the two-body
calculation at intermediate and high densities. The inclusion of the
Urbana-type TBF changes the equilibrium point
substantially, shifting it to
\begin{equation}
\rho_{\mathrm{sat}}=0.1710~\mathrm{fm}^{-3},
\qquad
E_{\mathrm{sat}}=-15.6434~\mathrm{MeV}.
\end{equation}
Thus, the TBF removes a considerable part of the
overbinding and moves the saturation density toward the empirical
nuclear-matter region. This behavior is consistent with the repulsive
character of the Urbana contribution at densities around and above
saturation. The introduction of $\Lambda$ hyperons modifies this balance in a
systematic manner. For both interaction schemes, increasing
$Y_{\Lambda}$ shifts the equilibrium point toward higher total baryon
densities and produces a more deeply bound minimum. In the two-body
calculation, the saturation density increases from
$0.3273~\mathrm{fm}^{-3}$ at $Y_{\Lambda}=0$ to
$0.3775~\mathrm{fm}^{-3}$ at $Y_{\Lambda}=0.1$ and
$0.4093~\mathrm{fm}^{-3}$ at $Y_{\Lambda}=0.2$. The corresponding
saturation energies change from $-23.3729~\mathrm{MeV}$ to
$-26.0172~\mathrm{MeV}$ and $-26.0877~\mathrm{MeV}$, respectively.
A similar trend is found after inclusion of the TBF,
although the equilibrium densities remain considerably lower:
\begin{equation}
\rho_{\mathrm{sat}}
=
0.1710,\ 0.2046,\ 0.2402~\mathrm{fm}^{-3}
\end{equation}
for $Y_{\Lambda}=0$, $0.1$, and $0.2$, respectively. This upward displacement of the equilibrium density with increasing
$Y_{\Lambda}$ can be understood from two related effects. First, the
replacement of nucleons by $\Lambda$ hyperons reduces the nucleonic
Fermi pressure and modifies the balance among the $NN$, $N\Lambda$,
and $\Lambda\Lambda$ correlation energies. Second, the Urbana-type
TBF acts only in the nucleonic sector. Consequently, at
fixed total baryon density, its repulsive contribution becomes less
effective when the nucleon fraction is reduced. The dependence is not
expected to follow a simple power of $1-Y_{\Lambda}$ because all
correlation functions and interaction contributions are determined
self-consistently. Nevertheless, the results clearly show a progressive
dilution of the nucleonic three-body repulsion as the $\Lambda$
fraction increases. This behavior provides a direct microscopic illustration of the
competition between hyperonic softening and three-body stiffening.
The nucleonic three-body interaction suppresses the excessive binding
and shifts the equilibrium state to lower density, whereas the presence
of $\Lambda$ hyperons weakens this repulsive mechanism and favors
equilibrium at a higher total density. The effect is already substantial
at $Y_{\Lambda}=0.1$ and becomes more pronounced at
$Y_{\Lambda}=0.2$. Therefore, although the Urbana-type force strongly
improves the saturation properties of purely nucleonic matter, its
ability to stiffen the system is progressively reduced as part of the
nucleonic component is replaced by hyperons. 

The incompressibility exhibits an interesting trend. In the two-body
calculation, $K_{\rm sat}$ increases from $370.9$~MeV at
$Y_{\Lambda}=0$ to $433.8$ and $444.9$~MeV at
$Y_{\Lambda}=0.1$ and $0.2$, respectively. With the inclusion of the
nucleonic TBF, the corresponding values are $304.6$,
$333.5$, and $380.3$~MeV. Since $K_{\rm sat}$ measures the curvature
of the energy per baryon around the equilibrium point, these results
indicate that matter with a finite prescribed $\Lambda$ fraction
becomes locally more resistant to compression around its own
composition-dependent saturation density. This local increase in
incompressibility does not, however, imply that the corresponding
hyperonic EOS is necessarily stiffer at suprasaturation densities.
The saturation density itself shifts with $Y_{\Lambda}$, so that the
quoted values of $K_{\rm sat}$ characterize different equilibrium
states. Moreover, away from saturation, the density dependence of the EOS is
increasingly governed by higher-order coefficients, such as
$Q_{\rm sat}$, and ultimately by the resulting pressure and its density
dependence. Consequently, a hyperonic EOS may
possess a larger local curvature around its shifted minimum while
remaining softer than the corresponding nucleonic EOS in the
high-density regime. 

The skewness coefficient also changes appreciably when the TBF is included. In particular, the increasingly negative values of
$Q_{\mathrm{sat}}$ indicate a stronger asymmetry of the energy curve
between the sub-saturation and supra-saturation sides of the minimum.
Since $Q_{\mathrm{sat}}$ depends on a third density derivative, it is
more sensitive to the fitting interval than
$\rho_{\mathrm{sat}}$, $E_{\mathrm{sat}}$, and
$K_{\mathrm{sat}}$. This is especially relevant for the
$Y_{\Lambda}=0.2$ calculation, for which the local fit gives
$Q_{\mathrm{sat}}\simeq-471~\mathrm{MeV}$ with a noticeably larger
fit uncertainty. 

The generalized symmetry-energy coefficients provide complementary
information about the response of the nucleonic subsystem to isospin
asymmetry. In the two-body calculation, the symmetry energy at
saturation decreases from $J=38.01~\mathrm{MeV}$
at $Y_{\Lambda}=0$ to $37.04~\mathrm{MeV}$ and
$33.85~\mathrm{MeV}$ at $Y_{\Lambda}=0.1$ and $0.2$,
respectively. The decrease reflects the modification of the nucleonic
isovector response when part of the baryonic density is carried by
isoscalar $\Lambda$ hyperons. After inclusion of the TBF, $J$ remains close to
$31~\mathrm{MeV}$ over the entire range of $Y_{\Lambda}$ considered.
The comparatively weak variation of $J$ suggests that the Urbana force
stabilizes the magnitude of the quadratic isospin response, in Eq.~\eqref{eq:isospin_expansion}, near the
composition-dependent equilibrium density. However, the slope
parameter increases from $L=79.1~\mathrm{MeV}$ at $Y_{\Lambda}=0$
to $81.0~\mathrm{MeV}$ and $83.0~\mathrm{MeV}$ at
$Y_{\Lambda}=0.1$ and $0.2$, respectively. The TBF
therefore has a more visible influence on the density dependence of
the symmetry energy than on its value at the equilibrium point.

The negative values of $K_{\mathrm{sym}}$ indicate a downward curvature
of the quadratic symmetry coefficient in the vicinity of the
composition-dependent saturation density. Its magnitude remains of
order $60$--$70~\mathrm{MeV}$ in the calculations including the
TBF. It should be emphasized that the coefficients obtained at
$Y_{\Lambda}>0$ are generalized bulk parameters of hypernuclear matter
at fixed $\Lambda$ fraction. They should not be identified directly
with the empirical saturation and symmetry-energy parameters of
ordinary nuclear matter. Nevertheless, their systematic evolution with
$Y_{\Lambda}$ provides a useful quantitative measure of how hyperons
modify both the isoscalar and isovector responses of the medium. Overall, the results demonstrate that the Urbana-type nucleonic
TBF is essential for correcting the excessive binding and
high saturation density generated by the two-body calculation.
However, because this repulsion acts only among nucleons, its relative
importance decreases as the $\Lambda$ fraction grows. The remaining
shift toward higher equilibrium density and stronger binding at finite
$Y_{\Lambda}$ indicates that nucleonic three-body repulsion alone
cannot completely compensate for the hyperonic softening mechanism.
This observation further motivates future investigations of repulsive
hyperonic three-body interactions, such as $NN\Lambda$ or
$N\Lambda\Lambda$ contributions, particularly when the model is
extended to beta-stable matter and NS densities.

\begin{table}[H]
\caption{Comparison of the LOCVY saturation and symmetry-energy
coefficients for symmetric nuclear matter with current empirical
and phenomenological constraints. The LOCVY results correspond to
$Y_{\Lambda}=0$ and $\delta_N=0$. The constraints on
$\rho_{\mathrm{sat}}$, $E_{\mathrm{sat}}$, $J$, and $L$ are the
$95\%$ credible intervals reported in Ref.~\cite{Drischler2024}.
The quoted constraint on $K_{\mathrm{sat}}$ is based on analyses
of giant monopole resonances adopted in Ref.~\cite{Tsang2024}.}
\label{tab:empirical_saturation}
\centering
\begin{tabular}{cccc}
\toprule
Quantity &
LOCV 2BF &
LOCV 2BF+TBF &
Reference constraint \\
\midrule
$\rho_{\mathrm{sat}}\;(\mathrm{fm}^{-3})$
& $0.3273$
& $0.1710$
& $0.157\pm0.010$ \\

$E_{\mathrm{sat}}\;(\mathrm{MeV})$
& $-23.3729$
& $-15.6434$
& $-15.97\pm0.40$ \\

$K_{\mathrm{sat}}\;(\mathrm{MeV})$
& $370.9$
& $304.6$
& $230\pm30$ \\

$Q_{\mathrm{sat}}\;(\mathrm{MeV})$
& $-72$
& $-233$
& poorly constrained \\

$J\;(\mathrm{MeV})$
& $38.01$
& $31.17$
& $32.0\pm1.1$ \\

$L\;(\mathrm{MeV})$
& $74.4$
& $79.1$
& $52.6\pm8.1$ \\

$K_{\mathrm{sym}}\;(\mathrm{MeV})$
& $-40$
& $-57$
& poorly constrained \\
\bottomrule
\end{tabular}
\end{table}

In Table \ref{tab:empirical_saturation}, we compare the saturation properties and symmetry-energy coefficients of symmetric matter obtained using the LOCVY method with empirical and phenomenological values. The comparison shows that the inclusion of the nucleonic TBF strongly improves the saturation density and saturation energy, while the resulting saturation density remains slightly above its inferred $95\%$ credible interval. The TBF also brings the symmetry-energy coefficient $J$ into agreement with the current reference range. Nevertheless, the calculated values of $K_{\mathrm{sat}}$ and $L$ remain above their respective reference ranges, indicating a relatively strong resistance of symmetric matter to compression and a relatively rapid increase of the symmetry energy with density in the vicinity of saturation.

\subsection{Microscopic decomposition of the energy}
\label{subsec:decomposition}

\begin{figure}[tbp]
    \centering

    \includegraphics[width=0.70\textwidth]{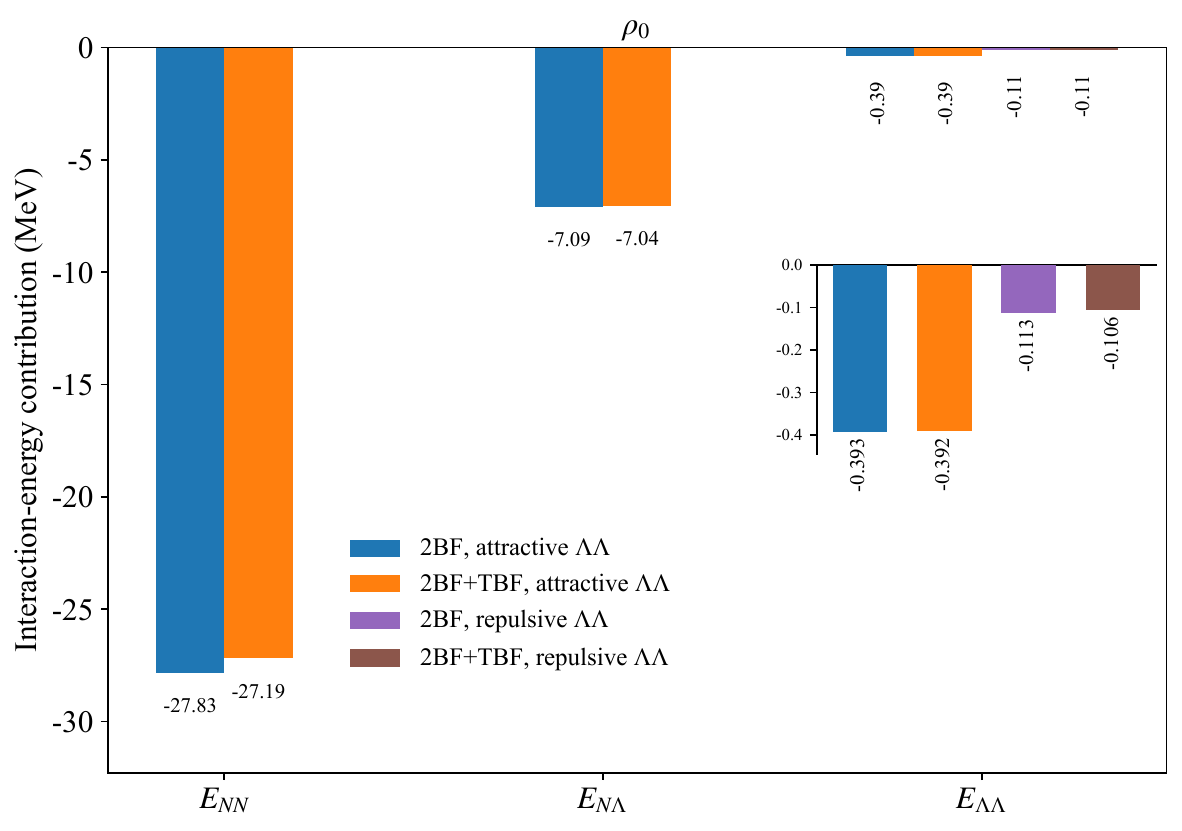}

    \includegraphics[width=0.70\textwidth]{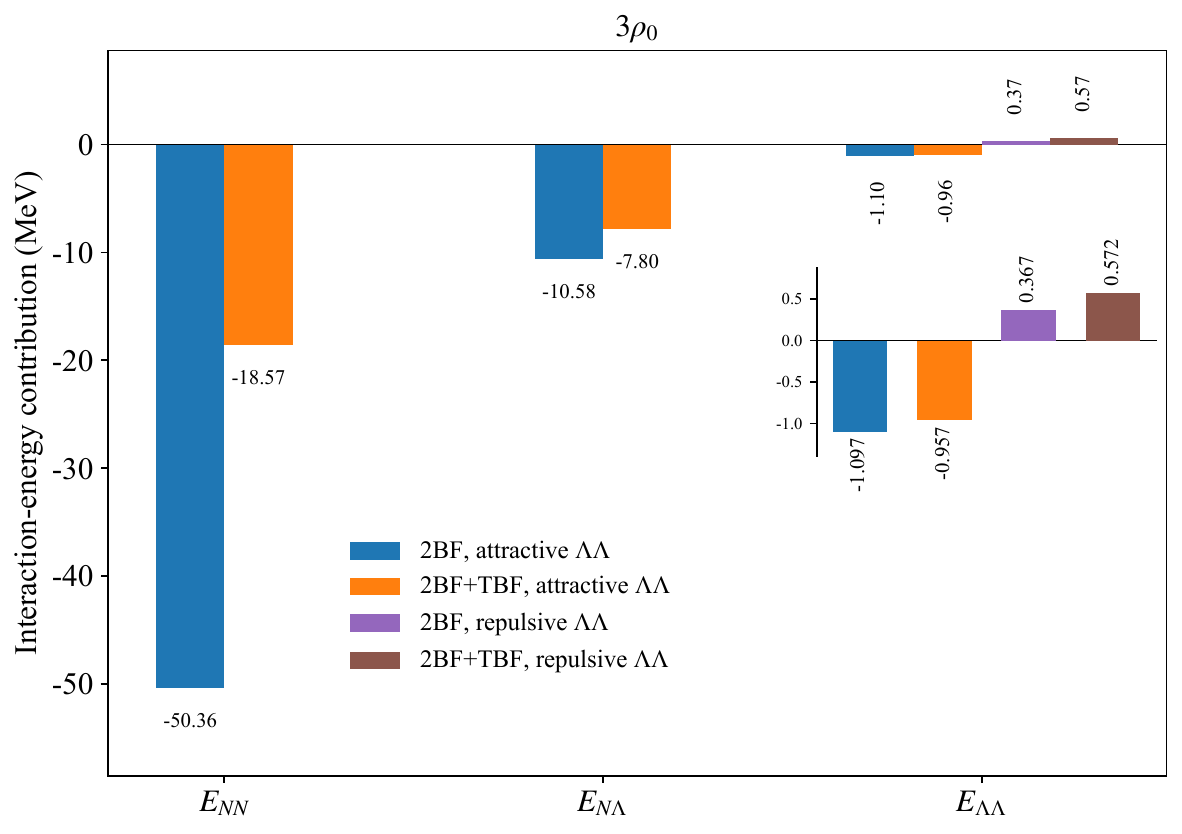}
    \vspace{0.3cm}
    \includegraphics[width=0.70\textwidth]{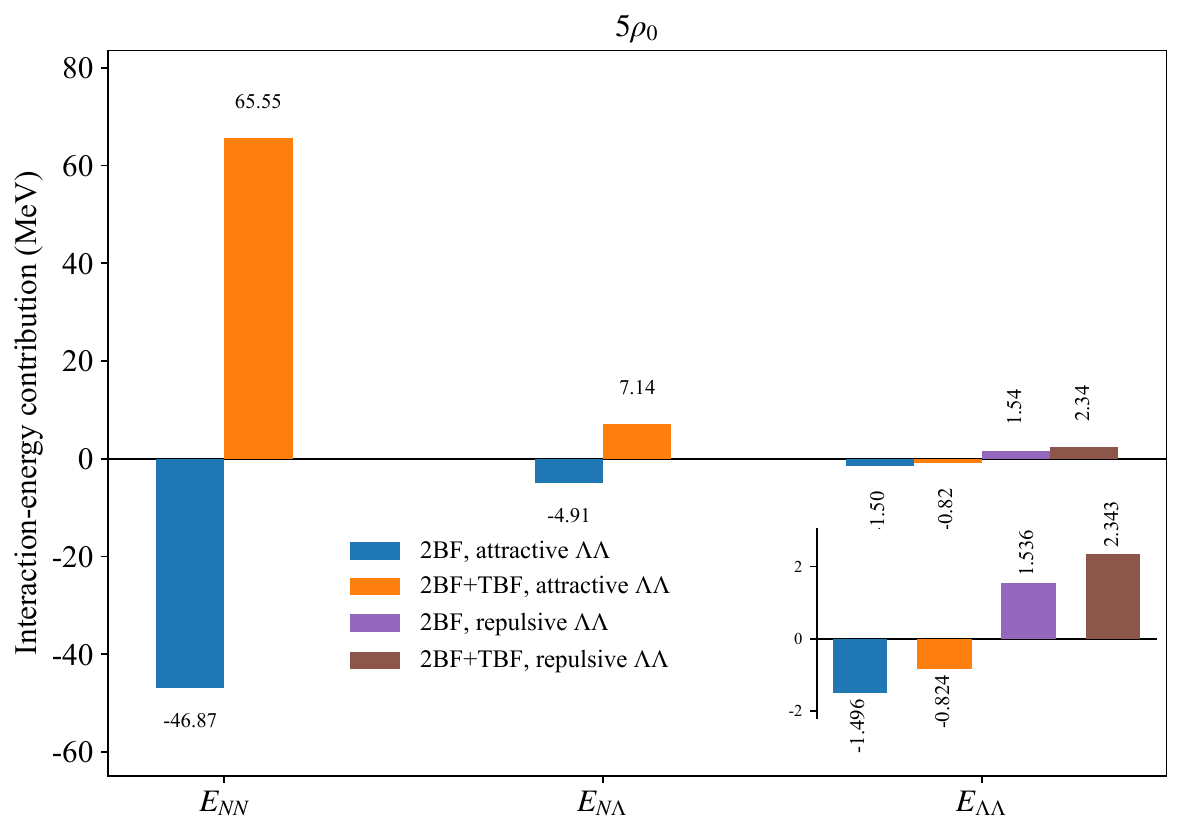}

    \caption{Decomposition of the interaction energy per baryon into the
nucleon--nucleon, nucleon--$\Lambda$, and $\Lambda\Lambda$ contributions
at three representative baryon densities.
The channel contributions are defined as
$E_{NN}=E_{nn}+E_{pp}+E_{np}$, $E_{N\Lambda}=E_{n\Lambda}+E_{p\Lambda}$ , and
$E_{\Lambda\Lambda}$. For the $NN$ and $N\Lambda$ sectors, the
blue and orange bars denote the 2BF and 2BF+TBF calculations,
respectively, using the attractive prescription for the
$\Lambda\Lambda$ interaction. For the $\Lambda\Lambda$ sector, the
attractive and repulsive prescriptions are both displayed, with the
same color convention used consistently in the main panels and the
insets. The insets magnify the comparatively small
$E_{\Lambda\Lambda}$ contribution. All results are shown for matter
with a symmetric nucleonic component and $Y_{\Lambda}=0.2$.}
    \label{fig:decomposition}
\end{figure}

Figure~\ref{fig:decomposition} illustrates the microscopic
decomposition of the interaction energy per baryon into the
nucleon--nucleon, nucleon--$\Lambda$, and $\Lambda\Lambda$ sectors at
$\rho_B=\rho_0$, $3\rho_0$, and $5\rho_0$ for matter with a symmetric
nucleonic component and $Y_{\Lambda}=0.2$. The comparison separates the
effects of the two-body interactions from those obtained after including
the Urbana-type nucleonic TBF. For the $NN$ and
$N\Lambda$ sectors, the results correspond to the attractive
$\Lambda\Lambda$ prescription, while both the attractive and repulsive
odd-state $\Lambda\Lambda$ prescriptions are shown for the
$\Lambda\Lambda$ contribution.

At $\rho_B=\rho_0$, the interaction energy is dominated by the attractive
$NN$ contribution, followed by the smaller $N\Lambda$ contribution.
Including the TBF changes $E_{NN}$ only moderately, from
approximately $-27.83$ to $-27.19~\mathrm{MeV}$, while its effect on
$E_{N\Lambda}$ is negligible at this density. The
$\Lambda\Lambda$ contribution is considerably smaller than the other
two terms. It remains attractive for both prescriptions, although its
magnitude is larger for the attractive prescription,
$E_{\Lambda\Lambda}\simeq-0.39~\mathrm{MeV}$, than for the repulsive
one, for which $E_{\Lambda\Lambda}\simeq-0.11~\mathrm{MeV}$.

The influence of the nucleonic TBF becomes much stronger
at $3\rho_0$. In the $NN$ sector, the interaction energy changes from
$-50.36~\mathrm{MeV}$ in the two-body calculation to
$-18.57~\mathrm{MeV}$ after the inclusion of UIX, demonstrating a
substantial reduction of the net attraction. The $N\Lambda$
contribution also becomes less attractive, changing from
$-10.58$ to $-7.80~\mathrm{MeV}$. Although UIX acts directly only among
nucleons, its inclusion modifies the $N\Lambda$ and
$\Lambda\Lambda$ contributions indirectly through the self-consistent
readjustment of the LOCVY correlation functions. The
$\Lambda\Lambda$ contribution nevertheless remains much smaller than
the $NN$ and $N\Lambda$ terms. At this density, the attractive
prescription gives a negative contribution of approximately
$-1~\mathrm{MeV}$, whereas the repulsive prescription produces a small
positive contribution.

The dominance of the nucleonic TBF is most evident at
$5\rho_0$. The $NN$ contribution changes from
$-46.87~\mathrm{MeV}$ in the two-body calculation to a strongly
repulsive value of $65.55~\mathrm{MeV}$ when UIX is included. The
$N\Lambda$ contribution undergoes a similar, although considerably
smaller, change, from $-4.91$ to $7.14~\mathrm{MeV}$. These results show
that the high-density increase of the energy is controlled primarily
by the nucleonic sector and is therefore consistent with the strong
stiffening of the EOS generated by UIX. By comparison, the
$\Lambda\Lambda$ term remains subleading: it is attractive for the
attractive prescription, changing from approximately
$-1.50$ to $-0.82~\mathrm{MeV}$, and repulsive for the repulsive
prescription, increasing from approximately $1.54$ to
$2.34~\mathrm{MeV}$.

Overall, the decomposition demonstrates that the Urbana-type
TBF has only a modest effect near saturation density but
becomes the dominant source of repulsion at several times
$\rho_0$. Its principal effect appears in the $NN$ sector, while the
changes in the $N\Lambda$ and $\Lambda\Lambda$ contributions arise
indirectly from the self-consistent reorganization of the many-body
correlations. The insets further emphasize that, despite the visible
sensitivity of $E_{\Lambda\Lambda}$ to the adopted odd-state
prescription, its absolute contribution remains small compared with
the $NN$ and $N\Lambda$ sectors over the density range considered.

\subsection{Comparison with other variational results}
\label{subsec:LLuncertainty}

\begin{figure}[H]
\centering

{\includegraphics[width=0.74\linewidth]{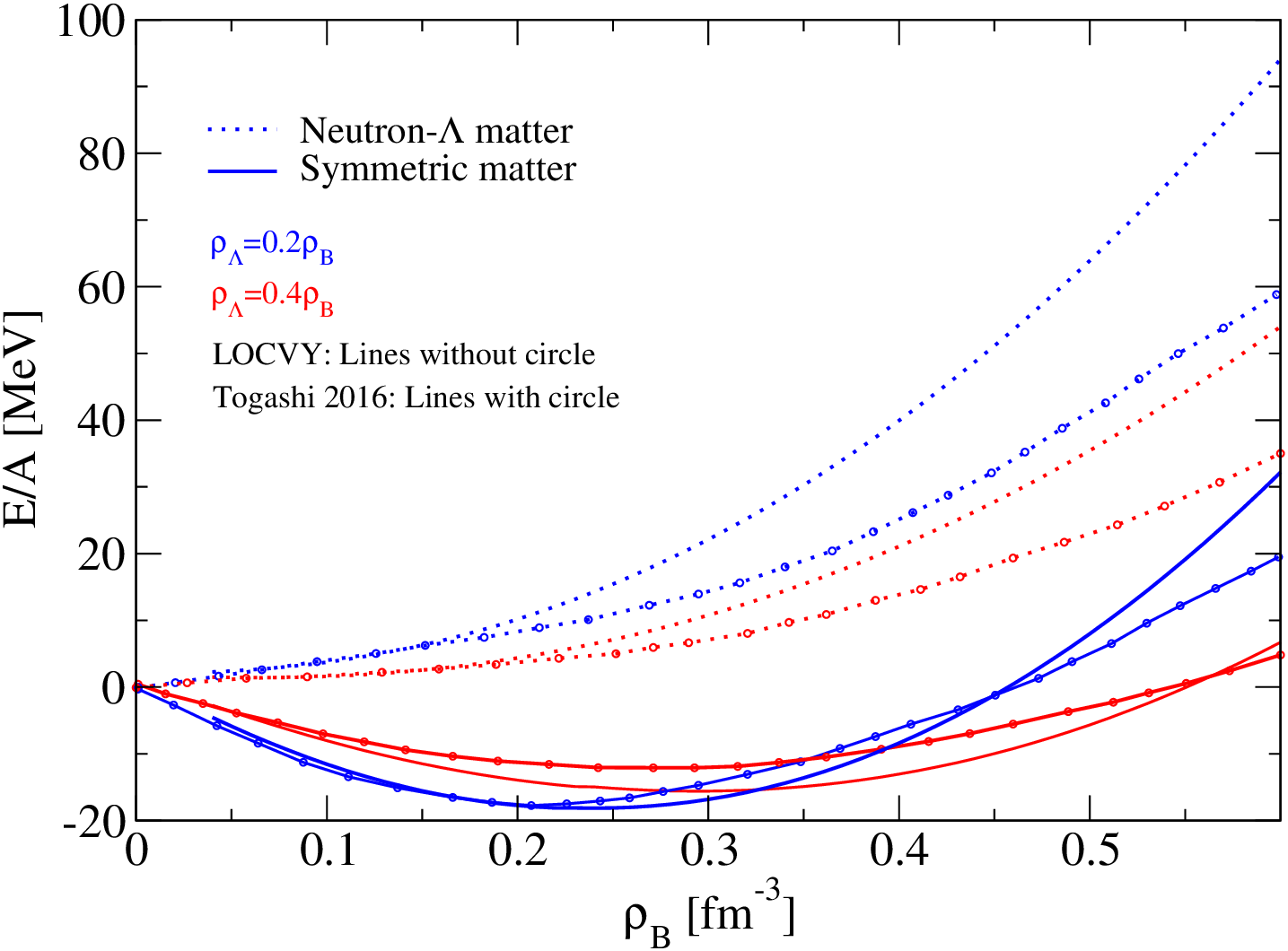}}

\caption{ The comparison of the EOS for hypernuclear matter
 calculated within the LOCVY method and the variational method
used in the work of Togashi et al. \cite{Togashi2016}. In both methods, the same two-body interactions for NN, NY, and YY are used, while the treatment of the TBF differs. The solid lines correspond to isospin symmetric matter while the dotted lines show the proton-free neutron-$\Lambda$
matter.
\label{fig:togashi}}
\end{figure}

Figure~\ref{fig:togashi} revisits the comparison with the
variational calculation of Togashi et al \cite{Togashi2016}. In our previous work
\cite{Shahrbaf2019}, the hyperonic EOS obtained within the LOCVY method did not include the UIX
TBF, whereas this contribution was already incorporated in the calculation
of Togashi et al. Consequently, part of the difference between
the two sets of results originated from the UIX TBF. In the present comparison, the same $NN$, $N\Lambda$, and
$\Lambda\Lambda$ interactions, together with UIX in the nucleonic sector,
are employed in both calculations. The resulting EOSs
are therefore considerably closer, particularly at low and intermediate
densities, showing that the inclusion of the same nucleonic TBF
removes a substantial part of the earlier discrepancy.

At higher densities, however, the LOCVY curves rise more rapidly than the
corresponding Togashi results for both the symmetric and neutron-rich
nucleonic components and for both values of $Y_{\Lambda}$. This behavior
indicates a stronger high-density repulsive response in the LOCVY
calculation and hence a comparatively stiffer EOS. Since the underlying
interactions are now matched, the remaining difference is more directly
associated with the distinct treatment of many-body correlations and
the variational constraints in the two approaches, rather than with a
mismatch in the adopted interactions.

%=================================================================
\section{Conclusions}
\label{sec:conclusions}
%=================================================================

We have extended the LOCVY description of homogeneous $n$-$p$-$\Lambda$ matter
by incorporating a UIX nucleonic TBF through a
correlation-averaged density-dependent two-nucleon interaction. The extension
preserves the $N\Lambda$ and $\Lambda\Lambda$ interactions of the earlier
two-body study, so that the difference between the two calculations isolates
the role of the improved nucleonic sector.

The analysis is organized around the competition between two mechanisms.
Replacing nucleons by $\Lambda$ hyperons generally lowers the high-density
energy and softens the EOS, whereas the UIX contribution becomes increasingly repulsive as the
nucleon density grows. For the fixed
fractions $Y_{\Lambda}=0$, $0.1$, and $0.2$, the calculated energy show that the nucleonic TBF partially compensates the hyperonic softening. The compensation is not expected
to be composition independent, because the direct UIX contribution is governed
by $\rho_N=(1-Y_{\Lambda})\rho_B$ and by the neutron--proton imbalance.

The analysis of the saturation properties provides a complementary bulk
perspective on the role of strangeness and many-body forces. For
$Y_{\Lambda}=0$, the inclusion of the nucleonic TBF shifts the
saturation point toward the empirical region of symmetric nuclear matter,
while finite prescribed $\Lambda$ fractions move the equilibrium point to
larger densities and modify the local curvature of the energy around
saturation. The corresponding symmetry-energy analysis, performed at fixed
$Y_{\Lambda}$, shows how the quadratic coefficient $S_2$ and its density
dependence evolve with increasing strangeness. The resulting $J$, $L$, and $K_{\rm sym}$ values for
$Y_{\Lambda}>0$ should therefore be interpreted as generalized
composition-dependent properties of hyperonic matter rather than as the
conventional empirical symmetry-energy parameters of ordinary nuclear
matter.

The decomposition of the interaction energy into $NN$, $N\Lambda$, and
$\Lambda\Lambda$ contributions provides a microscopic interpretation of
this bulk behavior. The resulting stiffening is not solely associated
with the direct contribution of the effective nucleonic three-body
interaction. Although UIX acts explicitly only in the nucleonic sector,
its inclusion modifies the self-consistent LOCVY solution and therefore
induces a rearrangement of the two-body correlation functions in all
baryonic channels. Consequently, the $N\Lambda$ contribution becomes
less attractive, and at sufficiently high density even repulsive, while
the $\Lambda\Lambda$ contribution is also shifted toward more repulsive
values for both the attractive and repulsive odd-state prescriptions.
These changes should be interpreted as medium-induced effects generated
by the self-consistent many-body treatment rather than as direct
three-body interactions in the hyperonic sector. Thus, in addition to
the dominant repulsion generated directly in the $NN$ sector, the
correlation rearrangement of the hyperonic channels further reinforces
the high-density stiffening of the EOS. The comparison of attractive and repulsive odd-state
$\Lambda\Lambda$ prescriptions also illustrates the uncertainty associated
with the poorly constrained hyperon--hyperon interaction. Its effect is
clearly visible in the $\Lambda\Lambda$ channel, but remains quantitatively
smaller than the high-density repulsion generated by the nucleonic
TBF.

The present fixed-composition calculation should be viewed as a microscopic
study of hyperonic many-body matter in its own right, rather than solely as
an intermediate step toward a complete neutron-star EOS. The resulting
density- and composition-dependent energies, saturation properties, and
channel-resolved $NN$, $N\Lambda$, and $\Lambda\Lambda$ contributions provide
information on the in-medium behavior of hyperonic interactions that is also
relevant to hypernuclear phenomenology and to the development of microscopic
descriptions of strange baryonic matter. In particular, the symmetric and
isospin-asymmetric limits considered here provide controlled reference cases
for exploring the role of strangeness in dense baryonic matter and may serve
as useful microscopic input for extensions toward the conditions encountered
in heavy-ion collisions, where finite-temperature and dynamical effects must
additionally be taken into account. A complementary extension to charge-neutral, $\beta$-equilibrated matter,
together with the inclusion of genuine hyperonic three-body forces,
will allow the consequences of the same microscopic framework for
NS structure to be assessed consistently.

%=================================================================
% MDPI back matter
%=================================================================
\vspace{6pt}

\authorcontributions{Conceptualization, M.Sh.; methodology, M.Sh.;
software, M.Sh.; validation, M.Sh.; formal analysis, M.Sh.;
investigation, M.Sh.; data curation, M.Sh.; writing--original draft
preparation, M.Sh.; writing--review and editing, M.Sh.;
visualization, M.Sh.; project administration, M.Sh.; funding
acquisition, M.Sh. The author has read and agreed to the published
version of the manuscript.}

\funding{M. Sh. has been supported by the National Science Center, Poland (NCN), under the SONATINA 7 Grant
No. 2023/48/C/ST2/00297.}

\institutionalreview{Not applicable.}

\informedconsent{Not applicable.}

\dataavailability{The numerical data supporting
the findings of this study will be made available in a public repository upon
publication.}

\acknowledgments{M. Sh. is deeply grateful to Makoto Oka for fruitful discussion. M. Sh. sincerely thanks Emiko Hiyama for the invitation to visit the RIKEN Nishina Center as a visiting
researcher and gratefully acknowledges the warm hospitality of the Few-body Systems in Physics Laboratory, RIKEN
Nishina Center, during her stay. During the preparation of this manuscript, the author used ChatGPT
(OpenAI) for assistance with language editing, improvement of clarity,
and refinement of the presentation of the manuscript. The author reviewed
and edited all outputs and takes full responsibility for the content of
the publication.}

\conflictsofinterest{The authors declare no conflicts of interest.}

\abbreviations{Abbreviations}{%
The following abbreviations are used in this manuscript:\\
\noindent
\begin{tabular}{@{}ll}
EOS & Equation of state\\
LOCV & Lowest-order constrained variational\\
LOCVY & Hyperonic lowest-order constrained variational\\
NS(s) & Neutron star(s)\\
TBF & Three-body force\\
UIX & Urbana IX\\
AV18 & Argonne $v_{18}$\\
2BF & Two-body force
\end{tabular}
}

%=================================================================
% References
%=================================================================

%\begin{adjustwidth}{-\extralength}{0cm}
\reftitle{References}

\isAPAandChicago{}{%

}
%\end{adjustwidth}

\end{document}